\documentclass[12pt]{article}
\usepackage{graphicx}
\usepackage{longtable}
\makeatletter
 
  \@addtoreset{equation}{section}
\makeatother
\def\beqar {\begin{eqnarray}}
\def\eeqar {\end{eqnarray}}
\def\beq {\begin{equation}}
\def\eeq {\end{equation}}

\def\F{{\cal F}}

\def\N{{\cal N}}

\def\al{\alpha}
\def\bt{\beta}
\def\del{\delta}

\def\ga{\gamma}
\def\Ga{\Gamma}

\def\ep{\epsilon}

\def\om{\omega}

\def\th{\theta}

\def\si{\sigma}

\def\zt{\zeta}

\def\d{\partial}

\def\Ad{{\dot A}}
\def\Bd{{\dot B}}

\def\bd{{\bar d}}

\def\bu{{\bar u}}

\def\hf{\frac{1}{2}}

\def\<{\langle}
\def\>{\rangle}

\def\Tr{{\rm Tr}}

\def\cp{{\bf CP}}

\begin{document}

\begin{titlepage}
\null\vspace{-62pt} \pagestyle{empty}
\begin{center}
\vspace{1.0truein}

{\Large\bf Holonomies of gauge fields in twistor space 7: \\
\vspace{.35cm}
\hspace{-.3cm}
an electroweak model} \\

\vspace{1.0in} {\sc Yasuhiro Abe} \\
\vskip .12in {\it Cereja Technology Co., Ltd.\\
1-13-14 Mukai-Bldg. 3F, Sekiguchi \\
Bunkyo-ku, Tokyo 112-0014, Japan } \\
\vskip .07in {\tt abe@cereja.co.jp}\\
\vspace{1.3in}
\centerline{\large\bf Abstract}
\end{center}
We consider massive deformation of $U(2)$ gauge bosons
in a recently developed holonomy formalism and propose a novel electroweak model.
The massive gauge bosons arise from massive deformation
of spinor momenta, which implies that the mass generation
is implemented by Lorentz symmetry breaking rather than
the spontaneous gauge symmetry breaking.
Following the notation of the holonomy formalism, we
interpret the weak hypercharge of a left-handed fermion doublet
as the reciprocal of the Knizhnik-Zamolodchikov (KZ) parameter
$\kappa = k + h^{\vee}$ where $k$ is the level number, fixed at $k = 1$,
and $h^\vee$ is the dual Coxeter number for $SU(2)_L$.
This leads to natural distinction between quarks and leptons
in terms of a weight for the representation of $SU(2)_L$.
Physical operators of the electroweak vector bosons and
fermions are defined by use of Grassmann variables.
Possible electroweak interactions are then determined by the evaluation of
Grassmann integrals.
We obtain a generating functional for the electroweak interaction vertices
and illustrate how to compute decay rates of the $Z$-boson into a pair of fermions.

\end{titlepage}
\pagestyle{plain} \setcounter{page}{2} 

\section{Introduction}

In a series of papers starting from \cite{Abe:2009kn},
we propose a novel framework, what we call the holonomy
formalism in twistor space, for the computation of gluon amplitudes.
The holonomy formalism has been developed in a way
of generalizing Nair's observation of the maximally helicity violating (MHV) amplitudes
that the MHV amplitudes can be interpreted as current correlators of
a Wess-Zumino-Witten (WZW) model suitably defined in supertwistor space \cite{Nair:1988bq}.
One of the technical, and probably conceptual, advantages of the holonomy formalism is
that it encompasses functional realization \cite{Abe:2004ep}
of the Cachazo-Svrcek-Witten (CSW) rules \cite{Cachazo:2004kj}
for tree amplitudes as well as for one-loop amplitudes \cite{Abe:2011af}.

In recent papers \cite{Abe:2012en,Abe:hol06}
we formulate S-matrix functionals for scattering amplitudes
of gluons and a pair of massive scalar/fermion particles
in the context of the holonomy formalism.
As a natural extension, in the present paper, we consider
incorporation of massive gauge bosons into the same framework.

Studies of electroweak phenomenology in terms of helicity
amplitudes have been carried out long before the recent
twistor-space based developments on the scattering amplitudes;
see, {\it e.g.}, \cite{Bern:1996ka,Dixon:1998py}.
For earlier studies on massive treatment of the spinor-helicity formalism,
see also \cite{Dittmaier:1998nn}.
More recently, applications of the CSW rules to the Higgs mechanism
are investigated in \cite{Dixon:2004za,Badger:2004ty,Buchta:2010qr}.
For other twistor-space inspired approaches to the standard electroweak model, see
\cite{Bern:2004ba,Badger:2005zh,Badger:2005jv}.

From a theoretical point of view, however, it is not
very clear why one can a priori apply the Higgs mechanism to the spinor-helicity formalism.
For example, the idea of ``on-shell constructibility'' for the construction of massive amplitudes
\cite{Cohen:2010mi,Boels:2011zz} implies that massive particles arise from massive deformations
of spinor momenta and supersymmetric Ward identities.
Thus it does not necessarily require the Higgs mechanism to generate massive particles.
The on-shell constructibility method has been applicable for massive scalars
\cite{Kiermaier:2011cr,Elvang:2011ub} as well as for massive fermions
\cite{Schwinn:2008fm,Huang:2012gs}, and there are no apparent obstacles that prevent
it from extending to massive vector bosons.
Indeed, in \cite{Boels:2011zz}, the so-called  maximally spin violating (MSV) amplitudes,
the massive-boson version of the MHV amplitudes, are proposed
by use of the on-shell constructibility method.

The absence of mass for a gauge field is generally required by
both gauge and Lorentz symmetries.
Thus, in a gauge-invariant framework such as the holonomy formalism,
masses are to be generated by Lorentz symmetry breaking at least effectively.
Notice that this is reminiscent of the fact that
a thermal mass term for gluons can be described by
a gauge-invariant WZW action \cite{Nair:1994fe,Alexanian:1995rp}.
(For a review of this result, see \cite{Nair:1994gr,Nair:2005iw}.
For related recent developments, see also \cite{Agarwal:2008rr,Agarwal:2009gb,Agarwal:2011gx}.)
Our construction of a novel electroweak model is partly motivated by these studies.

This paper is organized as follows.
In section 2 we first review basic ingredients of the holonomy formalism,
focusing on mass generation prescriptions for scalars and fermions.
In section 3 we apply these results to massive deformation of the electroweak gauge bosons.
We define physical operators of the electroweak bosons and
the fundamental fermions, {\it i.e.}, quarks and leptons.
Following these definitions, in section 4, we write down the electroweak
currents and Yukawa-type interaction terms in the holonomy formalism.
We then present a generating functional for the electroweak interaction vertices.
In section 5, using these results, we carry out practical calculations.
As a simple example, we compute decay rates of the $Z$-boson into a pair of fermions.
Lastly, in the concluding section, we briefly review our construction and
discuss positive and negative aspects of our formalism in comparison to the
standard electroweak model.

\section{Mass generation prescriptions in the holonomy formalism}

\noindent
\underline{Physical operators in the holonomy formalim}

In the holonomy formalism all the physical information is
embedded in the creation operators.
For gluons and their superpartners, the creation operators
are generically expressed as \cite{Abe:2009kn}:
\beq
    a_{i}^{(h_i )} (x , \th) \, = \, \left. \int d \mu(p_i) \,
    a_{i}^{(h_i )} (\xi_i^\al ) \, e^{i x \cdot p_i }
    \right|_{\xi^\al_i = \th_A^\al u^A}
    \label{2-1}
\eeq
where $h_i (= 0, \pm \hf, \pm 1)$ denotes the helicity
of the $i$-th particle ($i = 1,2, \cdots ,n $).
$p_i^\mu$ ($\mu = 0,1,2,3$) represents
the null momentum. In terms of the two-component spinor momentum
$u_i^A$ ($A = 1,2)$, the four-dimensional null momentum is expressed as
\beq
    p_{i}^{A \Ad} \, = \, ( \si_\mu )^{A \Ad} p_i^\mu
    \, \equiv \, u_i^A \bu_i^\Ad
    \label{2-2}
\eeq
where $\si^\mu = ({\bf 1}, \si^i )$, with ${\bf 1}$ and $\si^i$ ($i = 1,2,3$)
being the $2 \times 2$ identity matrix and the $2 \times 2$ Pauli matrices, respectively.
$\bu_i^\Ad$ ($\Ad = 1,2$) denotes the conjugate of $u_i^A$.
The spinor momenta are identical to the homogeneous coordinates on $\cp^1$.
Thus these are scale invariant. Besides, these are also invariant
under $U(1)$ phase transformations since the null momentum (\ref{2-2})
is invariant under such transformations.

The creation operator (\ref{2-1}) is defined on the
four-dimensional $\N=4$ chiral superspace whose coordinates
are represented by $x_\mu$ and $\th^\al_A$ $(\al = 1,2,3,4)$.
It is useful to introduce the Grassmann variable
\beq
    \xi_i^\al ~ = ~ \th^\al_A \, u_i^A
    \label{2-3}
\eeq
as shown in (\ref{2-1}).
Together with the twistor space condition
\beq
    v_{i \, \Ad} ~ = ~ x_{A \Ad} u_i^A \, ,
    \label{2-4}
\eeq
the variables $( u_i^A , v_{i \, \Ad} , \xi_i^\al )$ define the homogeneous
coordinates of the supertwistor space $\cp^{3|4}$.

In the spinor-helicity formalism the helicity of a particle can be defined as
\beq
    h_{i} = 1 - \hf  u_{i}^{A} \frac{\d}{\d u_{i}^{A}} \, .
    \label{2-5}
\eeq
This means that the helicity is determined by the number of $\xi_i^\al$'s
in the physical operator $a_{i}^{(h_i )} (\xi_i^\al )$.
For $\N=4$ extended supersymmetry, the number of $\xi_i^\al$'s
ranges from 0 to 4. Thus a full set of the operators
$a_{i}^{(h_i )} (\xi_i^\al )$ can be parametrized as
\beqar
    \nonumber
    a_{i}^{(+)} (\xi_i) &=& a_{i}^{(+)} \, ,
    \\
    \nonumber
    a_{i}^{\left( + \frac{1}{2} \right)} (\xi_i) &=& \xi_{i}^{\al}
    \, a_{i \, \al}^{ \left( + \frac{1}{2} \right)} \, ,
    \\
    a_{i}^{(0)} (\xi_i) &=& \hf \xi_{i}^{\al} \xi_{i}^{\bt} \, a_{i \, \al \bt}^{(0)}
    \, ,
    \label{2-6}
    \\
    \nonumber
    a_{i}^{\left( - \frac{1}{2} \right)} (\xi_i) &=&
    \frac{1}{3!} \xi_{i}^{\al}\xi_{i}^{\bt}\xi_{i}^{\ga}
    \ep_{\al \bt \ga \del} \, a_{i}^{ \left( - \frac{1}{2} \right) \, \del}
    \, ,
    \\
    \nonumber
    a_{i}^{(-)} (\xi_i) &=& \xi_{i}^{1} \xi_{i}^{2} \xi_{i}^{3} \xi_{i}^{4} \, a_{i}^{(-)} \, .
\eeqar
Notice that these operators correspond to massless particles
of the assigned helicity $h_i = (0, \pm \hf , \pm 1)$.
In the holonomy formalism the essence of Nair' superamplitude method
\cite{Nair:1988bq} lies in the above parametrization.

\noindent
\underline{Lorentz and gauge transformations in the holonomy formalim}

In the holonomy formalism physical variables are given by the spinor momenta.
Lorentz transformations of these momenta are expressed as
\beq
    u_i^A ~ \longrightarrow ~ u_{i}^{\prime \, A} \, = \,  ( g_u \, u_i )^A
    \, , ~~~
    \bu_i^\Ad ~ \longrightarrow ~ {\bu}_{i}^{\prime \, \Ad} \, = \, ( g_\bu \, \bu_i )^{\Ad}
    \label{2-7}
\eeq
where $g_u$ and $g_\bu$ denote the $(2 \times 2)$ matrix representation of
the $SL(2 ,{\bf C})$ algebra. The four-dimensional Lorentz symmetry
is then written as  $SL(2 ,{\bf C})_{u} \times SL(2 ,{\bf C})_{\bu}$.

On the other hand, color degrees of freedom for the creation operator
(\ref{2-1}) are embedded in the non-supersymmetric operator
$a_{i}^{( h_i )}$ in (\ref{2-6}). To be explicit, this is realized as
\beq
    a_{i}^{( h_i )} \, = \, t^{c_i}  a_{i}^{( h_i ) c_i}
    \label{2-8}
\eeq
where $t^{c_i}$ denotes the generator of the $U(N)$ gauge group.
A holonomy operator then becomes a single-trace operator in terms
of the color factor and, hence, it is invariant
under gauge transformations.

\noindent
\underline{Mass generation prescriptions for scalars and fermions: the $\xi\zt$-prescription}

We now review how to carry out off-shell continuation of Nair's superamplitude method
for scalars and fermions in the holonomy formalism.
To begin with, we define scalar products of $u^A$'s or $\bu_\Ad$'s:
\beq
    u_i \cdot u_j \, = \, (u_i \, u_j) \, = \,  \ep_{AB} u_{i}^{A}u_{j}^{B} \,
    \equiv \, ( i \, j ) \, , ~~~
    \bu_i \cdot \bu_j \, = \, [\bu_i \, \bu_j]  \, =  \,
    \ep_{\Ad \Bd} \bu_{i}^{\Ad} \bu_{j}^{\Bd} \, \equiv [ i \, j ]
    \label{2-9}
\eeq
where $\ep_{AB}$ and $\ep_{\Ad \Bd}$ are the rank-2 Levi-Civita tensors.
These products are invariant under the corresponding $SL(2,{\bf C})$ or
the two-dimensional Lorentz transformations.
These products are zero when $i$ and $j$ are identical.
In what follows, we can assume $1 \le i < j \le n$ without loss of generality.

In the massive spinor-helicity formalism \cite{Dittmaier:1998nn},
the spinor momenta undergo the massive deformation:
\beq
    u^A ~ \longrightarrow ~
    \widehat{u}^A \, = \,  u^A + \frac{m}{( u \eta ) } \eta^A
    \, , ~~
    \bu^\Ad ~ \longrightarrow ~
    \widehat{\bu}^\Ad \, = \,  \bu^\Ad + \frac{m}{[ \bu \bar{\eta} ] } \bar{\eta}^\Ad
    \label{2-10}
\eeq
where $\eta^A$ is a reference null spinor, with $\bar{\eta}$ being its conjugate.
$m$ denotes the mass of the deformed spinor momenta
$\widehat{u}^A$ and $\widehat{\bu}^\Ad$.
The corresponding massive four-momentum is defined as
$\widehat{p}^{A \Ad} = \widehat{u}^A \widehat{\bu}^\Ad$.

We then introduce another set of supertwistor variables
$( w^A , \pi_\Ad , \zt^\al )$ such that the supertwistor conditions
\beq
    \pi_\Ad \, = \, x_{\Ad A} w^A \, = \, x_{\Ad A} \frac{m}{( u \eta ) } \eta^A
    \, , ~~~
    \zt^\al \, = \, \th_{A}^{\al} w^A
    \, = \, \th_{A}^{\al} \frac{m}{( u \eta ) } \eta^A
    \label{2-11}
\eeq
are satisfied ($\al = 1,2,3,4)$.

Using the new Grassmann variable $\zt^\al_i = \th_{A}^{\al}
\frac{m}{( u_i \eta_i ) } \eta_i^A$, the off-shell continuation of the scalar operator
$a_{i}^{(0)} ( \xi_i )$ can be defined as \cite{Abe:2012en}:
\beq
    a_{i}^{(0)} ( \xi_i ) ~ \longrightarrow ~
    a_{i}^{(0)} ( \xi_i , \zt_i ) ~ = ~
    \xi_{i}^{1} \xi_{i}^{2} \xi_{i}^{3} \zt_{i}^{4} \, a_{i}^{(0)} \, .
    \label{2-12}
\eeq
The number of homogeneities in $u_i$'s remains the same in
the off-shell continuation (\ref{2-12}). Thus we can naturally interpret
$a_{i}^{(0)} ( \xi_i , \zt_i )$ as massive scalar operators.
We shall call this deformation the $\xi\zt$-prescription
to clarify our use of the ``massive'' Grassmann variable
\beq
    \zt^\al_i \, = \, \th_{A}^{\al} \frac{m}{( u_i \eta_i ) } \eta_i^A
    \, .
    \label{2-13}
\eeq
Notice that massive deformation of scalar operators are made
for a {\it pair} of scalar particles. We shall then specify
the numbering index to $i = 1, n$ in (\ref{2-12}).

Similarly, we find that the fermionic off-shell continuation can
be carried out by \cite{Abe:hol06}:
\beqar
    a_{Ri}^{\left( + \frac{1}{2} \right)} (\xi_i) ~ \longrightarrow ~
    a_{Ri}^{\left( + \frac{1}{2} \right)} ( \xi_i ) & = &
    \xi_{i}^{\al}  \, a_{Ri \, \al}^{ \left( + \frac{1}{2} \right)} \, ,
    \label{2-14} \\
    a_{Li}^{\left( - \frac{1}{2} \right)} (\xi_i) ~ \longrightarrow ~
    a_{Li}^{\left( - \frac{1}{2} \right)} ( \xi_i ) & = &
    \frac{1}{3!} \ep_{\al \bt \ga \del} \xi_{i}^{\al}\xi_{i}^{\bt}\xi_{i}^{\ga}
     \, a_{Li}^{ \left( - \frac{1}{2} \right)\, \del} \, ,
    \label{2-15} \\
    \bar{a}_{Li}^{\left( + \frac{1}{2} \right)} (\xi_i) ~ \longrightarrow ~
    \bar{a}_{Li}^{\left( + \frac{1}{2} \right)} ( \xi_i , \zt_i ) & = &
    \frac{1}{3!} \ep_{\al \bt \ga \del} \xi_i^\al \xi_i^\bt \zt_i^\ga
    \, \bar{a}_{Li}^{(+ \frac{1}{2} ) \del} \, ,
    \label{2-16} \\
    \bar{a}_{Ri}^{\left( - \frac{1}{2} \right)} (\xi_i) ~ \longrightarrow ~
    \bar{a}_{Ri}^{\left( - \frac{1}{2} \right)} ( \xi_i , \zt_i ) & = &
    \frac{1}{4} \xi_i^1 \xi_i^2 \xi_i^3 \xi_i^4 \zt_i^\al
    \, \bar{a}_{Ri \, \al}^{(- \frac{1}{2} ) }
    \label{2-17}
\eeqar
where $i = 1, n$. $\psi_{Li}$ and $\psi_{Ri}$ correspond to
the two-component Weyl spinors  with helicity $-\hf$ and $+ \hf$, respectively.
Notice that
the off-shell continuation by means of the $\xi\zt$-prescription is made only
for the conjugate fermions $\bar{\psi}_{Li}$ and $\bar{\psi}_{Ri}$ while
the unbar fermions $\psi_{Li}$ and $\psi_{Ri}$ remain on-shell.

For scalars and fermions, the chiral superspace representation of the massive operators
can be expressed as
\beqar
    \widehat{a}_{i}^{( 0 )} (x, \th)  & = &
    \left. \int d \mu ( \widehat{p}_i ) ~ a_{i}^{(0)} ( \xi_i , \zt_i )
    ~  e^{ i x_\mu \widehat{p}_i^\mu }
    \right|_{ \xi_{i}^{\al} = \th_{A}^{\al} u_i^A ,
    \, \zt_{i}^{\al} = \th_{A}^{\al} w_i^A  }
    \label{2-18} \\
    \bar{a}_{L1}^{\left( + \frac{1}{2} \right)}(x, \th)  & = &
    \left. \int d \mu ( \widehat{p}_1 ) ~ \bar{a}_{L1}^{\left( + \frac{1}{2} \right)} ( \xi_1 , \zt_1 )
    ~  e^{ i x_\mu \widehat{p}_1^\mu }
    \right|_{ \xi_{1}^{\al} = \th_{A}^{\al} u_1^A ,
    \, \zt_{1}^{\al} = \th_{A}^{\al} w_1^A  }
    \label{2-19} \\
    \bar{a}_{R1}^{\left( - \frac{1}{2} \right)}(x, \th)  & = &
    \left. \int d \mu ( \widehat{p}_1 ) ~ \bar{a}_{R1}^{\left( - \frac{1}{2} \right)} ( \xi_1 , \zt_1 )
    ~  e^{ i x_\mu \widehat{p}_1^\mu }
    \right|_{ \xi_{1}^{\al} = \th_{A}^{\al} u_1^A ,
    \, \zt_{1}^{\al} = \th_{A}^{\al} w_1^A  }
    \label{2-20} \\
    a_{L n}^{\left( - \frac{1}{2} \right)} (x, \th)  & = &
    \left. \int d \mu ( \widehat{p}_n ) ~ a_{L n}^{\left( - \frac{1}{2} \right)}  ( \xi_n )
    ~  e^{ i x_\mu \widehat{p}_n^\mu }
    \right|_{ \xi_{n}^{\al} = \th_{A}^{\al} u_n^A }
    \label{2-21} \\
    a_{R n}^{\left( + \frac{1}{2} \right)} (x, \th)  & = &
    \left. \int d \mu ( \widehat{p}_n ) ~ a_{R n}^{\left( + \frac{1}{2} \right)} ( \xi_n )
    ~  e^{ i x_\mu \widehat{p}_n^\mu }
    \right|_{ \xi_{n}^{\al} = \th_{A}^{\al} u_n^A }
    \label{2-22}
\eeqar
where
\beqar
    \widehat{p}_i^\mu & = & p_{i}^{\mu} + \frac{m^2}{2 (p_i \cdot \eta_i )} \eta_i^\mu
    \, ,
    \label{2-23} \\
    w_i^A & = &   \frac{m}{( u_i \eta_i ) } \eta_i^A
    \, .
    \label{2-24}
\eeqar
In the above expressions we specify the numbering indices of the massive fermion operators.
The massive four-momenta $\widehat{p}_1^\mu $ and $\widehat{p}_n^\mu$ are parametrized as
\beq
    \widehat{p}_1^\mu \, = \, p_1^\mu + \frac{m^2}{2 (p_1 \cdot p_n )} p_n^\mu
    \, , ~~~
    \widehat{p}_n^\mu \, = \, p_n^\mu + \frac{m^2}{2 (p_n \cdot p_1 )} p_1^\mu
    \, .
    \label{2-25}
\eeq
This means that we choose the reference null-vectors for the massive four-momenta as
\beq
    \eta_1^\mu = p_n^\mu \, , ~~~~ \eta_n^\mu = p_1^\mu \, .
    \label{2-26}
\eeq

\section{Mass generation prescriptions for the electroweak vector bosons}

Having reviewed the mass generation prescriptions for scalars and fermions,
in this section we consider application of the above results to gauge bosons.

\noindent
\underline{Peculiarities of the $U(2)_{L {\rm y}}$ gauge group}

From here on, we consider the electroweak gauge group
$U(2)_{L {\rm y}} = SU(2)_{L} \times U(1)_{\rm y}$.
In the holonoy formalism the physical operators (\ref{2-6}) are
holomorphic in terms of the spinor momenta.
As discussed in (\ref{2-7}),  the Lorentz symmetry of the holomorphic spinor momenta is
given by $SL(2 ,{\bf C})_{u}$, which is interpreted as
a complexification of $SU(2)_{u}$, {\it i.e.}, $SL(2, {\bf C})_{u} = SU(2)_{u}^{\bf C}$.
The spinor momenta is also invariant under phase transformations,
$u^A \rightarrow e^{i\th} u^A$ where $\th$ denotes the generator of $U(1)_u$.
This means that the ``holomorphic'' part of the Loretnz symmetry includes
the $U(2)_u$ group.

Now, we consider the symmetry breaking of $U(2)_u$
after the massive deformation of the spinor momenta,
$u^A \rightarrow  \widehat{u}^A = u^A + \frac{m}{( u \eta ) } \eta^A$.
Since the reference null spinor $\eta^A$ behaves like the spinor momentum,
it also preserves the $U(2)_u$ symmetry.
Under the phase transformation
$u^A \rightarrow e^{i\th} u^A$, the massive spinor momentum
becomes $\widehat{u}^A \rightarrow e^{i\th} u^A + e^{- i \th}
\frac{m}{( u \eta ) } \eta^A$.
Thus the $U(1)_u$ is broken upon the massive deformation.
On the other hand, the overall phase factor $U(1)_{\hat u}$
should be preserved since the off-shell four-momentum
$\widehat{p}^{A \Ad} = \widehat{u}^A \widehat{\bu}^\Ad$ is
invariant under the phase transformations
of $\widehat{u}^A \rightarrow e^{i \th^\prime} \widehat{u}^A$ and
$\widehat{\bu}^\Ad \rightarrow e^{- i \th^\prime} \widehat{\bu}^\Ad$
where $\th^\prime$ denotes the generator of $U(1)_{\hat u}$.

{\it
The original holomorphic Lorentz symmetry $U(2)_u$ for the massless spinor momentum
is therefore broken down to $U(2)_u \rightarrow SU(2)_u \times U(1)_{\hat u}$
due to the massive deformation of the spinor momentum.
}

We can understand this mechanism the other way around.
The physical operators $a_{i}^{( h_i )} ( \xi_i )$ are
is proportional to the Grassmann variables $\xi_i^\al = u_i^A \th_A^\al$,
while, as shown in (\ref{2-8}), the non-supersymmetric operator is given by
$a_{i}^{( h_i )}  = t^{c_i}  a_{i}^{( h_i ) c_i}$
where $t^{c_i}$ denotes the generator of the $U(2)_{L {rm y}}$ group.
Actions of the gauge transformation $U(2)_{L {\rm y}}$ and
those of the holomorphic Lorentz transformation $U(2)_{u}$
on the operators $a_{i}^{( h_i )} ( \xi_i )$
are therefore functionally equivalent.
In the context of symmetry breaking, this observation implies that
mass generation can be caused by the Lorentz symmetry breaking
rather than the spontaneous breaking of the gauge symmetry.

Motivated by these considerations, we now apply
the $\xi\zt$-prescription to the electroweak vector bosons.

\noindent
\underline{The $\xi\zt$-prescription for the electroweak vector bosons}

The massless spin $\pm$ gauge bosons $a_{i}^{(\pm)} (\xi_i )$ are defined in (\ref{2-6}).
Naive application of the $\xi \zt$-prescription to $a_{i}^{(\pm)} (\xi_i )$ is given by
\beqar
    a_{i}^{(+)} ( \xi_i ) & \longrightarrow &
    a_{i}^{(+)} ( \xi_i , \zt_i ) ~ = ~
    \hf \xi_{i}^{\al} \zt_{i}^{\bt} \, a_{i \, \al\bt}^{(+)} \, ,
    \label{3-1} \\
    a_{i}^{(-)} ( \xi_i ) & \longrightarrow &
    a_{i}^{(-)} ( \xi_i , \zt_i ) ~ = ~
    \frac{1}{12} \ep_{\al \bt \ga \del}
    \xi_{i}^{1} \xi_{i}^{2} \xi_{i}^{3}\xi_{i}^{4}\xi_{i}^{\al} \zt_{i}^{\bt}
    \, a_{i}^{(-) \ga\del} \, .
    \label{3-2}
\eeqar
We then interpret $a_{i}^{( \pm )} ( \xi_i , \zt_i )$
as the operators of the spin $\pm$ massive gauge bosons.
To include the spin-0 massive gauge bosons, we need to take account of
the above mass generation of the electroweak bosons, which is
caused by the holomorphic Lorentz symmetry breaking,
$U(2)_u \rightarrow SU(2)_u \times U(1)_{\hat u}$, rather than the
conventional spontaneous gauge symmetry breaking
$U(2)_{L {\rm y}} \rightarrow SU(2)_L \times U(1)_{\rm y}$.
As mentioned above, the $U(2)_u$ transformations play the
same role as the $U(2)_{L {\rm y}}$ transformations
when acted on the physical operators.

We now denote the original {\it massless} $U(2)_{L {\rm y}}$ gauge boson
by $b_{i}^{(h)}$ and $c_{i}^{(h)}$ ($h= \pm$)
where the former and the latter specify
the $SU(2)_L$ and the $U(1)_{\rm y}$ parts of the gauge boson, respectively.
Making the color factors explicit and the numbering indices implicit, these are
written as
\beq
    b^{(h)} = t^\al b^{(h) \al} \, , ~~~
    c^{(h)} = t^0 c^{(h)0}
    \label{3-3}
\eeq
where $t^\al = \frac{\si^\al}{2}$ and $t^0 = \frac{\bf 1}{2}$,
with $\si^\al$ ($\al=1,2,3$) being the $2 \times 2$ Pauli matrices and
${\bf 1}$ being the $2 \times 2$ identity matrix, respectively.

After the massive deformation
$U(2)_u \rightarrow SU(2)_u \times U(1)_{\hat u}$,
the weak bosons and the photon are conventionally parametrized as
\beqar
    {\rm w}^{(h)\pm} &=& \frac{b^{(h)1} \mp i b^{(h)2} }{ \sqrt{2} } \, ,
    \label{3-4} \\
    {\rm z}^{(h)} &=& - \sin \th_W \, c^{(h)0} \, + \, \cos \th_W \, b^{(h)3}
    \, , \label{3-5} \\
    a_{\rm em}^{(h)} &=& \cos \th_W \, c^{(h)0} \, + \, \sin \th_W \, b^{(h)3}
    \label{3-6}
\eeqar
where $\th_W$ arises from the phase difference between
$U(1)_{u}$ and $U(1)_{\hat u}$ transformations.
This angle corresponds to the Weinberg angle in the standard electroweak
model and is defined as
\beq
    \tan \th_W \, = \, \frac{g^\prime}{g}
    \label{3-7}
\eeq
where $g$ and $g^\prime$ are free parameters characterizing
the coupling constants of $SU(2)_L$ and $U(1)_{\rm y}$ gauge theories, respectively.

The massive deformation is effective to the
$SU(2)_u$ part of the $U(2)_u = SU(2)_u \times U(1)_u$ group.
The $U(1)_u$ part breaks due to the massive deformation but the massive spinor momentum
$\widehat{u}^A$ should be invariant under the overall $U(1)_{\hat u}$ phase transformation.
The overall $U(1)$ invariance is irrelevant to the massive deformation;
it is an embedded symmetry in the definition of the four-momentum in terms of the spinor momenta.
After the massive deformation
$U(2)_u \rightarrow SU(2)_u \times U(1)_{\hat u}$, the $U(1)_{\hat u}$ part then
remains massless.
In terms of the operators in (\ref{3-4})-(\ref{3-6}),
this means that ${\rm w}^{(h)\pm}$ and ${\rm z}^{(h)}$ receive the $\xi \zt$-prescriptions:
\beqar
    {\rm w}^{(h)\pm} & \longrightarrow & {\rm w}^{(\hat{h})\pm} ( \xi , \zt ) \, ,
    \nonumber \\
    {\rm z}^{(h)} & \longrightarrow & {\rm z}^{(\hat{h})} ( \xi , \zt )
    \, ,
    \label{3-29} \\
    c^{(h)} (\xi ) & \longrightarrow & a_{\rm em}^{(h)} (\xi )
\eeqar
where $\hat{h}=0, \pm$ and $h = \pm$.
To be explicit, the physical operators for the $W^{\pm}$ and $Z$ bosons are parametrized as
\beqar
    {\rm w}^{(0) \pm} (\xi , \zt ) & = & \xi^1 \xi^2  \xi^3 \zt^4 \, {\rm w}^{(0)\pm}
    \, ,
    \nonumber \\
    {\rm w}^{(+) \pm} (\xi , \zt ) & = & \hf \xi^{\al} \zt^{\bt} \, {\rm w}_{\al\bt}^{(+) \pm}
    \, ,
    \nonumber \\
    {\rm w}^{(-) \pm} (\xi , \zt ) & = &
    \frac{1}{12} \ep_{\al \bt \ga \del}
    \xi^{1} \xi^{2} \xi^{3}\xi^{4}\xi^{\al} \zt^{\bt}
    \,  {\rm w}^{(-) \pm \, \ga\del}
    \, ,
    \label{3-30} \\
    {\rm z}^{(0)} (\xi , \zt ) & = & \xi^1 \xi^2  \xi^3 \zt^4 \, {\rm z}^{(0)}
    \, ,
    \nonumber \\
    {\rm z}^{(+)} (\xi , \zt ) & = & \hf \xi^{\al} \zt^{\bt} \, {\rm z}_{\al\bt}^{(+)}
    \, ,
    \nonumber \\
    {\rm z}^{(-)} (\xi , \zt ) & = &
    \frac{1}{12} \ep_{\al \bt \ga \del}
    \xi^{1} \xi^{2} \xi^{3}\xi^{4}\xi^{\al} \zt^{\bt}
    \,  {\rm z}^{(-) \, \ga\del}
    \nonumber
\eeqar
while the photon operators remain massless:
\beqar
    a_{\rm em}^{(+)} (\xi ) &=& a_{\rm em}^{(+)} \, ,
    \nonumber \\
    a_{\rm em}^{(-)} (\xi ) &=& \xi^{1} \xi^{2} \xi^{3} \xi^{4} \, a_{\rm em}^{(-)} \, .
    \label{3-31}
\eeqar
Remember that the Grassmann variables $\xi^\al$ and $\zt^\al$ ($\al = 1,2,3,4$) are defined by
\beq
    \xi^\al \, = \, u^A \th_A^\al \, , ~~~~
    \zt^\al \, = \, \frac{m}{(u \, \eta)} \eta^A \th^\al_A
    \label{3-32}
\eeq
where we omit the numbering indices for the spinor momentum $u$
and the reference spinor $\eta$.

Notice that we begin with the massless $SU(2)_{L {\rm y}}$ gauge bosons
$( b^{(h)}, c^{(h)} )$. We then manually execute the massive deformation
of the gauge bosons by means of the $\xi \zt$-prescription and obtain
the electroweak gauge bosons $( {\rm w}^{(\hat{h})\pm}, {\rm z}^{(\hat{h})}, a_{\rm em}^{(h)})$
without introducing the Higgs mechanism.
The total number of states then increases by three, due to the production of
the massive spin-1 weak bosons $( {\rm w}^{(0)\pm}, {\rm z}^{(0)})$.

\noindent
\underline{Quarks, leptons and the weak hypercharges}

We introduce the coupling constants $g$ and $g^\prime$
in the definition of the Weinberg angle (\ref{3-7}).
In the holonomy formalism, however, the coupling constant is
determined by the reciprocal of the
KZ parameter $\kappa = k + h^{\vee}$ where $k$ is the level number, fixed at $k = 1$,
and $h^\vee$ is the dual Coxeter number for $SU(N)$, {\it i.e.},
\beq
    g_{hol} \, = \, \frac{1}{k + h^{\vee} } \, = \, \frac{1}{1 + N} \, .
    \label{3-8}
\eeq
See \cite{Abe:2009kn} for details of this relation.
Notice that we here use the fact that $h^\vee = N$
in the highest weight representation of the $SU(N)$ algebra.
Since the coupling constant $g_{hol}$ is built in the holonomy formalism,
we need to interpret it as some physical parameter even in an electroweak model.
In the following, we see that $g_{hol}$ can naturally be interpreted as
the weak hypercharge, a conserved quantum number in the standard electroweak model.

Recall that in the holonomy formalism massive fermions emerge as the
superpartners of gauge bosons, prescribed with the massive deformations
in (\ref{2-14})-(\ref{2-17}).
Thus, once the electroweak gauge group $U(2)_{L {\rm y}}$ is chosen,
the coupling constant $g_{hol} = \frac{1}{1 + h^{\vee}}$
would also be relevant to some quantum number for the quarks and leptons.

As mentioned above, the dual Coexter number for $SU(N)$ takes the value of $N$
in the highest weight representation. If we take the lowest weight representation
(or the negative root system), this number becomes $-N$.
Thus for $SU(2)$ we have two possibilities, $h^{\vee} = \pm 2$.
In the case of $SU(2)_L$, the coupling constant is then given by
\beq
    g_{hol} \, = \, \frac{1}{1 \pm 2} \, = \, \frac{1}{3} \, , \,  -1 \, .
    \label{3-9}
\eeq
These values correspond to
the weak hypercharges of the $SU(2)_L$ doublets for quarks and leptons, respectively,
{\it i.e.},
\beq
    {\rm y}( q_L^i ) = 1/3 \, , ~~~ {\rm y} ( l_L^i ) = -1
    \label{3-10}
\eeq
where $i$ denotes the generation $i = 1,2,3$ of the quark and the lepton doublets:
\beq
    q_L^i =
    \left(
    \begin{array}{c}
      u_L^i \\
      d_L^i
    \end{array}
    \right)  , ~~~
    l_L^i =
    \left(
    \begin{array}{c}
      \nu_L^i \\
      e_{L}^{i}
    \end{array}
    \right) .
    \label{3-11}
\eeq
At first glance, these results are a simple coincidence but
since $g_{hol}$ is relevant to some quantum number of fermions,
{\it it is natural to interpret that the quarks
are in the highest weight representation of $SU(2)$ (with $h^{\vee} = 2)$ while
the leptons are in the lowest weight representation (with $h^{\vee} = - 2)$.}
We do not have a satisfying explanation for this interpretation yet.
It is, however, intriguing that this interpretation provides mathematical
distinction between quarks and leptons. Physically, they are of course distinct;
the quarks couple with gluons but the leptons do not.
The above interpretation is thus expected to be useful in the incorporation of
QCD into the holonomy formalism.

Including the right-handed fermions, the full hypercharges
of fermions are expressed as
\beq
    \begin{array}{ll}
      {\rm y}( u_L ) = 1/3 \, , & ~ {\rm y}( \nu_L ) = -1 \, , \\
      {\rm y}( d_L ) = 1/3 \, , & ~ {\rm y}( e_L ) = -1 \, , \\
      {\rm y}( u_R ) = 4/3 \, , & ~ {\rm y}( \nu_R ) = 0 \, , \\
      {\rm y}( d_R ) = -2/3 \, , & ~ {\rm y}( e_R ) = -2
    \end{array}
    \label{3-12}
\eeq
where we omit the generation indices.
The right-handed fermions, {\it i.e.}, the $SU(2)$ singlets,
do not have hypercharges which are identical to the left-handed counterparts.
The left- and right-handed fermions are related to each other by
\beq
    \begin{array}{ll}
    {\rm y} ( u_R ) \, = \, {\rm y} ( u_L ) + 1 \, , &
    ~ {\rm y}( \nu_R ) \, = \, {\rm y} ( \nu_L ) + 1 \, , \\
    {\rm y} ( d_R ) \, = \, {\rm y} ( d_L ) - 1 \, , &
    ~ {\rm y}( e_R ) \, = \, {\rm y} ( e_L ) - 1 \, .
    \end{array}
    \label{3-13}
\eeq
These relations suggest that a right-handed fermion can be
interpreted as a composite of a left-handed fermion and a particle of
hypercharge $\pm 1$.

\noindent
\underline{Introduction of auxiliary Higgs-like scalars}

According to (\ref{3-8}), the particle of hypercharge $1$
would be characterized by $N=0$ and, hence, we can
interpret it as a scalar particle.
Its anti-particle then becomes the particle of hypercharge $-1$.
We shall denote these by $\phi$ and $\bar{\phi}$, respectively:
\beq
    {\rm y} ( \phi ) = 1 \, , ~~~~ {\rm y}( \bar{\phi} ) = -1 \, .
    \label{3-14}
\eeq
By use of these we can express the right-handed fermions as
\beq
    \begin{array}{ll}
    u_R \, \sim \, \phi \, u_L \,  , & ~~~ d_R \, \sim \, \bar{\phi}\, d_L \, , \\
    \nu_R \, \sim \, \phi \, \nu_L \,  , & ~~~ e_R \, \sim \, \bar{\phi}\, e_L \, .
    \end{array}
    \label{3-15}
\eeq
Since the hypercharge is additive, these relations are in agreement with those of (\ref{3-13}).
The above expression implies that the right-handed fermions can
be expressed as composite particles.

Introduction of $\phi$ and $\bar{\phi}$ is suitable for the explanation
of not only the hypercharge disparity in (\ref{3-13})
but also the discrepancies in the number of $\xi$'s in
the definitions of massive fermion operators (\ref{2-14})-(\ref{2-17}).
To be more concrete, we can define the pair of the scalar operators as
\beq
    \phi ( \xi )\, = \, \hf \xi^\al \xi^\bt \phi_{\al \bt} \, , ~~~
    \bar{\phi} ( \xi )\, = \, \hf  \xi^\al \xi^\bt  \bar{\phi}_{\al \bt}
    \label{3-16}
\eeq
where we omit the numbering indices.
The number of $\xi$'s for scalars is uniquely determined as shown in (\ref{2-6}).
Together with (\ref{3-15}), the above definitions are in accord with
the massive fermion operators of (\ref{2-14})-(\ref{2-17}).

Notice that $\phi$ and $\bar{\phi}$ are {\it massless} scalars.
Massive deformation of these can be carried out as
\beq
    \phi ( \xi ) ~ \longrightarrow ~
    \phi ( \xi , \zt )  ~ = ~
    \xi^{1} \xi^{2} \xi^{3} \zt^{4} \, \phi
    \label{3-17}
\eeq
and the same for $\bar{\phi}$.
These are analogous to the Higgs scalars in a sense
that they are necessary to generate mass for fermions.
In our framework, however, these are no longer $SU(2)$ doublets.
We have derived these scalars in such a way that is
consistent with the correspondence between
the coupling constant $g_{hol}$ and the weak hypercharge ${\rm y}$,
and also with our definitions of the massive fermion operators.

\noindent
\underline{Electroweak couplings}

In the standard model the electroweak gauge field is defined as
\beq
    a_{L {\rm y}}^{(h)}
    \, = \, g \, b^{(h)} \, + \, {\rm y} g^\prime \, c^{(h)}
    \, = \,
    g \left(
    b^{(h)} + {\rm y} \tan \th_W \, c^{(h)}
    \right) \, .
    \label{3-18}
\eeq
This definition holds after the massive deformation of
the operators $b^{(h)}$ and $c^{(h)}$ in (\ref{3-3}).
Thus, using (\ref{3-4})-(\ref{3-6}), (\ref{3-30}) and (\ref{3-31}),
the operators can be parametrized by ${\rm w}^{(h) \pm} (\xi, \zt)$,
${\rm z}^{(h)} (\xi , \zt )$ and $a_{\rm em}^{(h)} ( \xi )$
where we denote the spin/helicity by $h$ in either massive or massless operators.
The weak hypercharge ${\rm y}$ is built in the above definition.
Since in the standard model ${\rm y}$ relates to the electromagnetic coupling
constant $Q$ by
\beq
    Q \, = \, t_L^3 + \frac{{\rm y}}{2} \, ,
    \label{4-8}
\eeq
the definition (\ref{3-18}) implies that the electroweak bosons
couple to fermions. Indeed, in the standard model the electroweak
bosons couple with the charged and neutral currents,
each of which is represented by a pair of fermions.

The hypercharge ${\rm y}$ in (\ref{3-18}) is  then dependent on the fermions to be coupled with.
For the left-handed quarks, ${\rm y}( q_L^i ) = 1/3$,
the gauge field $a_{L {\rm y}}^{(h)}$ can be expressed as
\beq
    a_{L {\rm y} ( q_L^i )}^{(h)}
    \, = \,
    \left(
      \begin{array}{cc}
        \frac{2}{3} e a_{\rm em}^{(h)} + \left( \frac{1}{2} - \frac{2}{3} \sin^2 \th_W \right) g_z {\rm z}^{(h)}
        & g {\rm w}^{(h)+} \\
        g {\rm w}^{(h)-} & -\frac{1}{3}e a_{\rm em}^{(h)} +
        \left( -\frac{1}{2} + \frac{1}{3} \sin^2 \th_W \right) g_z {\rm z}^{(h)} \\
      \end{array}
    \right)
    \label{3-19}
\eeq
where
\beqar
    e &=& \frac{g g'}{\sqrt{g^2 + g'^2}} \, ,
    \label{3-20} \\
    g_z  &=& \frac{g}{\cos \th_W } ~ = ~ \sqrt{g^2 + g'^2 } \, .
    \label{3-21}
\eeqar
For the left-handed leptons, ${\rm y} ( l_L^i ) = -1$, we have
\beq
    a_{L {\rm y} ( l_L^i )}^{(h)}
    \, = \,
    \left(
      \begin{array}{cc}
        \frac{1}{2} g_z {\rm z}^{(h)} & g {\rm w}^{(h)+} \\
        g {\rm w}^{(h)-} & - e a_{\rm em}^{(h)} - \left( \frac{1}{2} - \sin^2 \th \right) g_z {\rm z}^{(h)} \\
      \end{array}
    \right) .
    \label{3-22}
\eeq

In the right-handed sector the electroweak gauge field is described solely by
$c^{(h)}$, {\it i.e.},
\beq
    a_{R {\rm y}}^{(h)}
    \, = \,   {\rm y} g^\prime \, c^{(h)}
    \, .
    \label{3-23}
\eeq
Substituting the values of ${\rm y}$'s in (\ref{3-12}), we can similarly
express the corresponding gauge fields as
\beqar
    a_{R {\rm y} ( u_R )}^{(h)}
    & = &
    \frac{2}{3} e a_{\rm em}^{(h)} - \frac{2}{3} \sin^2  \th_W g_z {\rm z}^{(h)} \, ,
    \label{3-24} \\
    a_{R {\rm y} ( d_R )}^{(h)}
    & = &
    - \frac{1}{3} e a_{\rm em}^{(h)} + \frac{1}{3} \sin^2  \th_W g_z {\rm z}^{(h)} \, ,
    \label{3-25} \\
    a_{R {\rm y} ( \nu_R )}^{(h)}
    & = &
    0 \, ,
    \label{3-26} \\
    a_{R {\rm y} ( e_R )}^{(h)}
    & = &
    -  e a_{\rm em}^{(h)} +  \sin^2  \th_W g_z {\rm z}^{(h)} \, .
    \label{3-27}
\eeqar

The coefficients of ${\rm w}^{(h)\pm}$, ${\rm z}^{(h)}$ and
$a_{\rm em}^{(h)}$ can be interpreted as the coupling constants
between the electroweak bosons and the electroweak currents.
In the next section we define the electroweak currents
and show the validity of these coefficients.

\section{Electroweak currents and Yukawa-type couplings}

\noindent
\underline{Electroweak currents}

Following the previous section, we consider the couplings between
the electroweak vector bosons and the electroweak charged/neutral currents.
The currents are in a bilinear form of the fermions.
The gauge fields are given by (\ref{3-19})-(\ref{3-27}), depending on
the type of fermions to couple with.

For simplicity we consider one generation/family model
in what follows, omitting the generation indices of quarks and leptons.
The interaction terms between electroweak bosons and the electroweak currents
can be expressed as
\beqar
    \Ga_{\rm int}^{(h)} &=&
    \bar{q}_L \, a_{L {\rm y}( q_L )}^{(h)} q_L
    \, + \,
    \bar{l}_L \, a_{L {\rm y}( l_L )}^{(h)} l_L
    \nonumber \\
    &&
    \!\!\!\!\!\!\! + ~
    \bar{u}_R \, a_{R {\rm y}( u_R )}^{(h)} u_R
    \, + \,
    \bar{d}_R \, a_{R {\rm y}( d_R )}^{(h)} d_R
    \, + \,
    \bar{\nu}_R \, a_{R {\rm y}( \nu_R )}^{(h)} \nu_R
    \, + \,
    \bar{e}_R \, a_{R {\rm y}( e_R )}^{(h)} e_R
    \label{4-1}
\eeqar
where
\beq
    \bar{q}_L =
    \left(
      \begin{array}{cc}
        \bu_L & \bd_L \\
      \end{array}
    \right)  , ~~~~
    \bar{l}_L =
    \left(
      \begin{array}{cc}
        \bar{\nu}_L & \bar{e}_L \\
      \end{array}
    \right) .
    \label{4-2}
\eeq
Using (\ref{3-18})-(\ref{3-27}), we find that the interaction terms
can be rewritten as
\beq
    \Ga_{\rm int}^{(h)} ~ = ~ e \, a_{\rm em}^{(h)} J^{\rm em}
    \, + \, g_z \, {\rm z}^{(h)} J^{0}
    \, + \,
    g \left( {\rm w}^{(h)+} J^{-} + {\rm w}^{(h)-} J^{+} \right)
    \label{4-3}
\eeq
where
\beqar
    J^{\rm em} & = &
    \frac{2}{3} (  \bu_L u_L +  \bu_R u_R ) - \frac{1}{3}
    ( \bd_L d_L + \bd_R d_R ) - ( \bar{e}_L e_L    + \bar{e}_R e_R ) \, ,
    \label{4-4} \\
    J^{0} & = & \hf ( \bu_L u_L - \bd_L d_L ) + \hf ( {\bar \nu}_{L} \nu_{L} - \bar{e}_L e_L )
    - \sin^2 \th_W \, J^{\rm em} \, ,
    \label{4-5} \\
    J^{-} & = & \bu_L d_L \, + \, \bar{\nu}_L e_L \, ,
    \label{4-6} \\
    J^{+} &=& \bd_L u_L \, + \, \bar{e}_L \nu_L \, .
    \label{4-7}
\eeqar
Notice that the coefficients in the electromagnetic current (\ref{4-4}) agree with
the electromagnetic coupling constants $Q =  t_L^3 + \frac{{\rm y}}{2}$ defined in (\ref{4-8}).

\noindent
\underline{Supersymmetrization and comparison to the standard electroweak model}

The electroweak currents (\ref{4-4})-(\ref{4-7}) are
the same in structure as those of the standard electroweak model
except that the latter are defined with the gamma matrices.
In our parametrization the physical operators
are labeled by the helicity/spin and the numbering index;
note that we have been making the numbering indices implicit in the above expressions.
Furthermore, a crucial difference is that in the holonomy formalism the physical
operators are dependent on the Grassmann variables and
we eventually carry out Grassmann integrals over them.

The number of the Grassmann variables attached to each fermion operator
is defined as (\ref{2-14})-(\ref{2-17}).
Consequently, we find that the number of the Grassmann variables for
any of the electroweak currents (\ref{4-4})-(\ref{4-7}) become six.
This means that these currents should couple with ${\rm w}^{(+) \pm} (\xi , \zt )$ or
${\rm z}^{(+)} (\xi , \zt )$, having two Grassmann variables,
otherwise the interaction terms vanish upon the evaluation of
the Grassmann integrals over $\th^\al_A$'s ($\al = 1,2,3,4$).
It is interesting that the above helicity-based arguments,
equipped with the Grassmann integrals, determine
the non-vanishing helicity configuration of the electroweak interactions.

Regarding the nature of the electroweak current,
there are a few important differences from the standard model.
First, the Grassmann integral of $\Ga_{\rm int}^{(h)}$ gives rise to
a fermion mass. Let $\bar{\psi}_{L/R} \psi_{L/R}$ be a generic
form of an electroweak current, then $m_\psi$, the mass of $\psi$,
is extracted out of the Grassmann integrals over $\th^\al_A$'s.

Second, owing to the saturation of the Grassmann variables,
the current-current interactions vanish upon the evaluation of the
Grassmann integrals. The arrowed vertices among the electroweak
bosons and the fermions are given by (\ref{4-3}) and
Yukawa-type interactions which we shall discuss in a moment.
The current-current interactions in the standard electroweak model
are derived in low-energy limits. Thus, the absence of these interactions
does not contradict the standard model since in the present case we
consider at energy levels which are compatible with the masses of the
weak gauge bosons.

Lastly, the electromagnetic coupling $a_{\rm em}^{(\pm)} (\xi ) J^{\rm em} (\xi , \zt)$
vanishes upon taking the Grassmann integrals because
the photon operators (\ref{3-31}) have either zero or four Grassmann variables.
This result is not derived from the standard model but is in agreement
with the result obtained in the study of the massive fermion amplitudes \cite{Abe:hol06}
in which we show that a gluon/photon should couple with
a pair of massive fermions of the form in either
$\bar{\psi}_{L} \psi_R$ or $\bar{\psi}_{R} \psi_L$.
The fermion mass $m_\psi$ also emerges upon the evaluation of the Grassmann
integrals for these Yukawa-type interaction terms.

{\it In the holonomy formalism fermion mass arises from supersymmetrization
of the physical operators and consequent evaluation of the Grassmann integrals,
rather than the introduction of Higgs fields or the Higgs mechanism.}

\noindent
\underline{Yukawa-type couplings in the holonomy formalism}

We now define the Yukawa-type couplings for quarks and leptons in the holonoy formalism.
We first consider quarks.
The up-quark mass terms arise from the bilinear forms $\bu_L u_R$ and $\bu_R u_L$.
The number of Grassmann variables attached to these are four and eight, respectively.
This means that in the electroweak model $\bu_R u_L$ couples only with
the positive-helicity photon $a_{\rm em}^{(+)}$.
On the other hand, $\bu_L u_R$ can couple with
the negative-helicity photon $a_{\rm em}^{(-)}$ and
the spin-0 weak bosons ${\rm w}^{(0)\pm}$, ${\rm z}^{(0)}$.

Since $u_L$ forms a doublet with $d_L$, the mass terms
relevant to $\bu_R u_L$ can naturally be calculated as
\beq
    \left(
      \begin{array}{cc}
        \bu_R & \bd_R \\
      \end{array}
    \right)
    a_{L {\rm y}(q_L )}^{(h)}
    \left(
      \begin{array}{c}
        u_L \\
        d_L \\
      \end{array}
    \right) ~
    \Longrightarrow ~
    \frac{2}{3}e \, a_{\rm em}^{(+)} \, \bu_R  u_L \, - \,
    \frac{1}{3}e \, a_{\rm em}^{(+)} \, \bd_R  d_L
    \label{4-9}
\eeq
where the right-arrow indicates that we show only the terms
that have eight Grassmann variables upon supersymmetrization, {\it i.e.}, those terms
that survive after the Grassmann integral.
The right-handed sector can also be calculated as
\beqar
    \bu_L a_{R {\rm y} (u_R )}^{(h)} u_R
    & \Longrightarrow &
    \frac{2}{3} e \, a_{\rm em}^{(-)} \, \bu_L u_R
    \, - \,
    \frac{2}{3} \sin^2 \th_W g_z {\rm z}^{(0)} \,  \bu_L u_R
    \, ,
    \label{4-10} \\
    \bd_L a_{R {\rm y} (d_R )}^{(h)} d_R
    & \Longrightarrow &
    - \frac{1}{3} e \, a_{\rm em}^{(-)} \, \bd_L d_R
    \, + \,
    \frac{1}{3} \sin^2 \th_W g_z {\rm z}^{(0)} \,  \bd_L d_R
    \, .
    \label{4-11}
\eeqar
As mentioned above $\bu_L u_R$ and $\bd_L d_R$
can interact with the spin-0 $Z$-boson ${\rm z}^{(0)}$.
Notice that the spin-0 $W^{\pm}$ bosons do not involve here
since these bosons do not enter in the definitions of
$a_{R {\rm y} (u_R )}^{(h)}$ and $a_{R {\rm y} (d_R )}^{(h)}$
in (\ref{3-24}) and (\ref{3-25}), respectively.

Similarly, lepton mass terms are described as
\beq
    \left(
      \begin{array}{cc}
        \bar{\nu}_R & \bar{e}_R \\
      \end{array}
    \right)
    a_{L {\rm y}(l_L )}^{(h)}
    \left(
      \begin{array}{c}
        \nu_L \\
        e_L \\
      \end{array}
    \right) ~
    \Longrightarrow ~
    - e \, a_{\rm em}^{(+)} \, \bar{e}_R  e_L
    \label{4-12}
\eeq
and
\beqar
    \bar{\nu}_L a_{R {\rm y} (\nu_R )}^{(h)} \nu_R
    & \Longrightarrow &
    0
    \, ,
    \label{4-13} \\
    \bar{e}_L a_{R {\rm y} (e_R )}^{(h)} e_R
    & \Longrightarrow &
    -  e \, a_{\rm em}^{(-)} \, \bar{e}_L e_R
    \, + \,
    \sin^2 \th_W g_z {\rm z}^{(0)} \,  \bar{e}_L e_R
    \, .
    \label{4-14}
\eeqar
There are no neutrino mass terms in the above.
This is due to our definitions of
$a_{L {\rm y}(l_L )}^{(h)}$, $a_{R {\rm y} (\nu_R )}^{(h)}$
and $a_{R {\rm y} (e_R )}^{(h)}$.
Notice that the neutrinos are defined as massive fermions,
satisfying (\ref{2-14})-(\ref{2-17}) from the beginning.
Thus lack of the neutrino mass terms does not necessarily
mean massless neutrinos as in the standard electroweak model.
In fact, in our framework the neutrino mass arises from the
term $g_z {\rm z}^{(+)} J^{0}$ in (\ref{4-3}).
At the present stage, however, it is not clear why neutrino
masses are experimentally so tiny in comparison to the other fundamental fermions.

Lastly, for completion of the argument, we express the Yukawa-type terms
in analogy with the electroweak interaction terms $\Ga_{\rm int}^{(h)}$ in (\ref{4-1}):
\beqar
    \Ga_{\rm Yuk}^{(h)}  &=&
    \left(
      \begin{array}{cc}
        \bu_R & \bd_R \\
      \end{array}
    \right)
    a_{L {\rm y}(q_L )}^{(h)}
    \left(
      \begin{array}{c}
        u_L \\
        d_L \\
      \end{array}
    \right)
    \, + \,
    \left(
      \begin{array}{cc}
        \bar{\nu}_R & \bar{e}_R \\
      \end{array}
    \right)
    a_{L {\rm y}(l_L )}^{(h)}
    \left(
      \begin{array}{c}
        \nu_L \\
        e_L \\
      \end{array}
    \right)
    \nonumber \\
    &&
    \!\!\!\!\!\!\! + ~
    \bar{u}_L \, a_{R {\rm y}( u_R )}^{(h)} u_R
    \, + \,
    \bar{d}_L \, a_{R {\rm y}( d_R )}^{(h)} d_R
    \, + \,
    \bar{\nu}_L \, a_{R {\rm y}( \nu_R )}^{(h)} \nu_R
    \, + \,
    \bar{e}_L \, a_{R {\rm y}( e_R )}^{(h)} e_R
    \, .
    \label{4-15}
\eeqar
Using the expressions in section 3, we can also describe $a_{L {\rm y}}^{(h)}$'s
and $a_{R {\rm y}}^{(h)}$'s in terms of the Weinberg angle $\th_W$
and the electroweak bosons, {\it i.e.},
$a_{\rm em}^{(h)}$, ${\rm w}^{{(h) \pm}}$, ${\rm z}^{(h)}$.
To be explicit, $\Ga_{\rm Yuk}^{(h)}$ can alternatively be expressed as
\beq
    \Ga_{\rm Yuk}^{(h)} ~ = ~ e \, a_{\rm em}^{(h)} J_{\rm Yuk}^{\rm em}
    \, + \, g_z \, {\rm z}^{(h)} J_{\rm Yuk}^{0}
    \, + \,
    g \left( {\rm w}^{(h)+} J_{\rm Yuk}^{-} + {\rm w}^{(h)-} J_{\rm Yuk}^{+} \right)
    \label{4-16}
\eeq
where
\beqar
    J_{\rm Yuk}^{\rm em} & = &
    \frac{2}{3} (  \bu_R u_L +  \bu_L u_R ) - \frac{1}{3}
    ( \bd_R d_L + \bd_L d_R ) - ( \bar{e}_R e_L    + \bar{e}_L e_R ) \, ,
    \label{4-17} \\
    J_{\rm Yuk}^{0} & = & \hf ( \bu_R u_L - \bd_R d_L ) + \hf ( {\bar \nu}_{R} \nu_{L} - \bar{e}_R e_L )
    - \sin^2 \th_W \, J_{\rm Yuk}^{\rm em} \, ,
    \label{4-18} \\
    J_{\rm Yuk}^{-} & = & \bu_R d_L \, + \, \bar{\nu}_R e_L \, ,
    \label{4-19} \\
    J_{\rm Yuk}^{+} &=& \bd_R u_L \, + \, \bar{e}_R \nu_L \, .
    \label{4-20}
\eeqar

\noindent
\underline{Summary}

To recapitulate our formalism, we now consider a holonomy operator of the electroweak model.
The holonomy operator that generates the massive fermion amplitudes,
{\it i.e.}, the amplitudes of gluons and a pair of massive fermions, is defined by \cite{Abe:hol06}
\beqar
    &&
    \Theta_{R, \ga}^{(B)_{\bar{\psi}\psi} } (u; x , \th)
    \nonumber \\
    &=& \exp \Biggl[ \,
    \sum_{r \ge 3}
    \sum_{ ( h_2 , h_3 , \cdots , h_{r-1} ) }
    g^{r-2} ( -1 )^{ h_2 h_3 \cdots h_{r-1} }
    \,
    \Tr \left(
    t^{c_2} t^{c_3} \cdots t^{c_{r-1}}
    \right)
    \frac{
    a_{2}^{(h_2)c_2} \otimes \cdots \otimes a_{r-1}^{(h_{r-1})c_{r-1}}
    }{
    (12)(23) \cdots (r-1 \, r)( r 1)
    }
    \nonumber \\
    &&
    ~~~~~~ \otimes \,
    \left(
    a_{Lr}^{\left( - \frac{1}{2} \right)} \otimes \bar{a}_{R1}^{\left( - \frac{1}{2} \right)}
    +
    a_{Rr}^{\left( + \frac{1}{2} \right)} \otimes \bar{a}_{L1}^{\left( + \frac{1}{2} \right)}
    +
    a_{Lr}^{\left( - \frac{1}{2} \right)} \otimes \bar{a}_{L1}^{\left( + \frac{1}{2} \right)}
    +
    a_{Rr}^{\left( + \frac{1}{2} \right)} \otimes \bar{a}_{R1}^{\left( - \frac{1}{2} \right)}
    \right) \,
    \Biggr]
    \label{4-21}
\eeqar
where $h_{i} = \pm = \pm 1$ ($i=2,3, \cdots, r-1$) denotes
the helicity of the $i$-th gluon.
Here the gluon operator $a_{i}^{(h_i)} = t^{c_i} a_{i}^{(h_i )c_i}$
is defined in the superspace representation $a_{i}^{(h_i)} (x, \th)$
as shown in (\ref{2-1}).
Similarly, the fermionic operators $a_{Lr}^{\left( - \frac{1}{2} \right)}$,
$\bar{a}_{R1}^{\left( - \frac{1}{2} \right)}$, etc. are defined in the
superspace representation as shown in (\ref{2-19})-(\ref{2-22}).

In terms of $\Theta_{R, \ga}^{(B)_{\bar{\psi}\psi} } (u; x , \th)$,
the generating functional for the massive fermion ultra-helicity-violating (UHV)
vertices, {\it i.e.}, the vertices of positive-helicity gluons and a pair of massive fermions,
is then expressed as
\beq
    \F_{\rm UHV}^{\rm (vertex)} \left[ a^{( \pm )c} ,
    \bar{a}_{L/R}^{\left( \pm \frac{1}{2} \right)} ,
    a_{L/R}^{\left( \mp \frac{1}{2} \right)}
    \right]
    \, = \,
    \exp \left[
    i \int d^4 x d^8 \th ~ \Theta_{R , \ga}^{(B)_{\bar{\psi} \psi} } (u; x, \th)
    \right] \, .
    \label{4-22}
\eeq
As studied in \cite{Abe:hol06}, building blocks of the massive
fermion amplitudes are given by three-point vertices.
The fact that the massive fermion amplitudes are decomposed into
the three-point vertices suggests that building blocks of
the electroweak interactions are given by the terms in (\ref{4-3})
and the Yukawa-type terms (\ref{4-16}).
In analogy of the form (\ref{4-22}), the generating functional
for these interaction vertices can then be expressed as
\beqar
    \F_{\rm int}^{\rm (vertex)} \left[
    a_{\rm EW}^{(h)} ,
    \bar{a}_{\rm EW}^{\left( \pm \frac{1}{2} \right)} ,
    a_{\rm EW}^{\left( \mp \frac{1}{2} \right)}
    \right]
    & = &
    \exp \left[
    i \int d^4 x d^8 \th ~ \sum_{h=\pm 1, 0} (-1)^{h}
    \frac{\Ga_{\rm EW}^{(h)} (u; x, \th )}{(12)(23)(31)}
    \right]
    \, ,
    \label{4-23} \\
    \Ga_{\rm EW} (u; x , \th)
    & = & \Ga_{\rm int}^{(h)} (u; x , \th) +  \Ga_{\rm Yuk}^{(h)} (u; x, \th) \, .
    \label{4-24}
\eeqar
where $a_{\rm EW}^{(h)}$ represents a set operators for the electroweak bosons and
$\left( \bar{a}_{\rm EW}^{\left( \pm \frac{1}{2} \right)},
a_{\rm EW}^{\left( \mp \frac{1}{2} \right)} \right)$ generically denote
operators of anti-fermions and fermions, with the handedness implicit.
As discussed in (\ref{4-21}) the argument $(u ; x, \th)$ means
that $\Ga_{\rm int}^{(h)}$ and $\Ga_{\rm Yuk}^{(h)}$
are defined in the superspace representation.
Notice that the numbering indices are implicit in $\Ga_{\rm EW} (u; x , \th)$;
as in the case of massive fermion vertices, the numbers
1, 2, 3 are assigned to $\bar{a}_{\rm EW}^{\left( \pm \frac{1}{2} \right)}$,
$a_{\rm EW}^{(h)}$ and $a_{\rm EW}^{\left( \mp \frac{1}{2} \right)}$, respectively.
{\it
Finally, we would like to emphasize again that the Grassmann integral in (\ref{4-23}) provides
essential features of our electrweak model. It is this Grassmann integral
that makes our model different from the standard one.
}

\section{$Z$-boson decay rates}

Having defined the generating functional for the electroweak interaction vertices
in the previous section, we now carry out some practical calculations.
In this section, as a simple example, we consider decay processes of the $Z$  boson
into a pair of fermions.

We first consider the decay rate of the $Z$-boson into a pair of
electric neutrinos, $Z \rightarrow \bar{\nu_e}\nu_e$.
(In the following we denote $\nu_e$ by $\nu$ as before since in the present paper
we are dealing with a one-family model.)
From $\Ga_{\rm EW} (u; x , \th)$ in (\ref{4-24}),
the non-vanishing interaction term of interest is given by
$g_z  {\rm z}^{(+)} \bar{\nu}_L \nu_L$. Its supersymmetric form
is then expressed as
\beqar
    &&
    \hf \, g_z  \, \bar{\nu}_{L}^{\left( + \hf \right)} (\xi_1 , \zt_1 ) \, {\rm z}^{(+)} (\xi_2 , \zt_2 )
    \, \nu_{L}^{\left( - \hf \right)} ( \xi_3 )
    \nonumber \\
    &=&
    \hf g_z \,
    \frac{1}{3!} \ep_{\al \bt \ga \del} \xi_1^\al \xi_1^\bt \zt_1^\ga
    \, \bar{\nu}_{L}^{\left( + \hf \right) \del}  \,
    \frac{1}{3!} \ep_{\al^\prime \bt^\prime \ga^\prime \del^\prime}
    \xi_{3}^{\al^\prime}\xi_{3}^{\bt^\prime}\xi_{3}^{\ga^\prime}
    \, \nu_{L}^{\left( - \hf \right) \del^\prime}
    \,
    \hf \xi_{2}^{\al^{\prime \prime}} \zt_{2}^{\bt^{\prime \prime}} \,
    {\rm z}_{\al^{\prime \prime} \bt^{\prime \prime} }^{(+)}
    \,
    \label{5-1}
\eeqar
where we use (\ref{2-15}), (\ref{2-16}) and (\ref{3-30}).
As before $\xi_i^\al$ is defined as $\xi_i^\al = u_i^A \th_A^\al$ while
the ``massive'' Grassmann variables $\zt_1^\al$, $\zt_2^\al$ are parametried as
\beqar
    \zt_1^\al &=& \frac{m_\nu}{ (u_1 \eta_1 )} \eta_1^A \th_A^\al \, = \,
    \frac{m_\nu}{(13)} \xi_3^\al \, ,
    \label{5-2} \\
    \zt_2^\al &=& \frac{m_Z}{ (u_2 \eta_2 )} \eta_2^A \th_A^\al
    \label{5-3}
\eeqar
where we use the specific choice of $\eta_{1} = u_3$ as before; see
(\ref{2-26}). On the other hand, the reference spinor $\eta_2$ for the $Z$-boson
is a priori unspecified.
The Grassmann integral of (\ref{5-1}) can be calculated as
\beqar
    &&
    \hf \, g_z  \, \int d^8 \th ~
    \bar{\nu}_{L}^{\left( + \hf \right)} (\xi_1 , \zt_1 ) \, {\rm z}^{(+)} (\xi_2 , \zt_2 )
    \, \nu_{L}^{\left( - \hf \right)} ( \xi_3 )
    \nonumber \\
    &=&
    \hf \, g_z \, m_\nu m_Z \, \frac{(13)(23)(3 \eta_2 )}{( 2 \eta_2 )}
    ~ \bar{\nu}_{L}^{\left( + \hf \right) \del}  \,
    {\rm z}_{\del \del^{\prime} }^{(+)} \,
    \nu_{L}^{\left( - \hf \right) \del^\prime} \, .
    \label{5-4}
\eeqar
Following the convention, we consider the decay rate in the momentum-space
representation, imposing the periodic boundary conditions for spatial directions.
\beqar
    \Ga_{hol} (Z \rightarrow \bar{\nu}\nu )
    & = &
    \int \frac{d^3 \widehat{p}_1}{(2 \pi )^3} \frac{d^3 \widehat{p}_3}{(2 \pi )^3} \,
    \frac{1}{2 \om_{\hat{p}_1} 2 \om_{\hat{p}_2} 2 \om_{\hat{p}_3} }
    \, (2 \pi )^4 \del^{(4)}( \widehat{p}_2 - \widehat{p}_1 - \widehat{p}_3 )
    \nonumber \\
    &&
    ~~~ \times
    \left|
     \hf \, g_z \, m_\nu m_Z \, \frac{(13)(23)(3 \eta_2 )}{( 2 \eta_2 )}
     \frac{1}{(12)(23)(31)}
    \right|^2
    \label{5-5}
\eeqar
The incoming off-shell four-momentum for the $Z$-boson is defined as
\beq
    \widehat{p}_2^\mu \, =  \, (  \om_{\hat{p}_2} , \vec{p}_2 )
    \, =  \,
    p_2^\mu + \frac{m_Z^2} {2 (p_2 \cdot \eta_2 )} \eta_2^\mu \, .
    \label{5-6}
\eeq
For the pair of neutrinos the outgoing off-shell four-momenta are given by
\beqar
    \widehat{p}_1^\mu &=& (  \om_{\hat{p}_1} , \vec{p}_1 )
    \, = \,
    p_1^\mu + \frac{m_\nu^2} {2 (p_1 \cdot p_3 )} p_3^\mu \, ,
    \label{5-7} \\
    \widehat{p}_3^\mu &=& (  \om_{\hat{p}_3} , \vec{p}_3 )
    \, = \,
    p_3^\mu + \frac{m_\nu^2} {2 (p_3 \cdot p_1 )} p_1^\mu \, .
    \label{5-8}
\eeqar

In the rest frame of the initial particle, we can parametrize the above quantities as
\beqar
    &&
    \widehat{p}_2^\mu = (m_Z , 0, 0, 0) \, , ~~
    p_2^\mu = \left( \frac{m_Z}{2} , 0, 0, \frac{m_Z}{2} \right) \, , ~~
    \eta_2^\mu = ( \eta_2^0 , 0 , 0, - \eta_2^0 ) \, ,
    \label{5-9} \\
    &&
    \widehat{p}_1^\mu = ( \om_{\hat{p}} , \vec{p} ) \, , ~~~~
    p_1^\mu = ( p^0_1 , \vec{p}_1 ) \, ,
    \label{5-10} \\
    &&
    \widehat{p}_2^\mu = ( \om_{\hat{p}} , - \vec{p} ) \, , ~~~~
    p_2^\mu = ( p^0_1 , - \vec{p}_1 ) \,
    \label{5-11}
\eeqar
where $\om_p = \sqrt{m_\nu^2 + \vec{p}^2 }$ and $(p^0_1 )^2 - | \vec{p}_1 |^2 = 0$.
Substituting these into (\ref{5-5}), the decay rate can be simplified as
\beqar
    \Ga_{hol} (Z \rightarrow \bar{\nu}\nu )
    & = &
    \frac{g_z^2 m_Z m_\nu^2 }{128 \pi^2} \int d^3 \vec{p_1}
    \frac{\del (m_Z - 2 \om_p ) }{ \om_p^2 }
    \, \left|
    \frac{ ( 3  \eta_2 ) }{ (2 \eta_2 ) (12) }
    \right|^2
    \nonumber \\
    &=&
    \frac{g_z^2 m_Z m_\nu^2 }{128 \pi^2} \,
    2 \pi \sqrt{1 - \left( \frac{2 m_\nu}{m_Z} \right)^2}
    \,  \frac{ p_3 \cdot \eta_2 }{2 ( p_2 \cdot \eta_2 ) ( p_1 \cdot p_2 ) }
    \nonumber \\
    &=&
    \frac{g_z^2 m_Z}{64 \pi}  \left( \frac{m_\nu}{m_Z} \right)^2
    \sqrt{1 - \left( \frac{2 m_\nu}{m_Z} \right)^2}
    \label{5-12}
\eeqar
where we use the relation
\beq
    | ( u u' ) |^2 \, = \,  (u u' ) [ \bu \bu' ] \, = \, 2 p^\mu p'_\mu \, = \,
    2 ( p \cdot p' ) \, .
    \label{5-13}
\eeq

In the standard electrweak model, the decay rate of $Z$-boson to a pair of fermions
at tree level (for a one-family model) is given by
\beq
    \Ga_{\rm SM} (Z \rightarrow \bar{f}f )
    \, = \,
    \frac{g_z^2 m_Z}{24 \pi} \left( a_f^2 + b_f^2 \right)
    \label{5-14}
\eeq
where
\beq
    a_f = t_{L f}^{3}  - Q_f \sin^2 \th_W
    \, , ~~~
    b_f = - Q_f \sin^2 \th_W \, .
    \label{5-15}
\eeq
For $f = \nu$, we have $a_\nu = \hf$ and $b_\nu = 0$, leading to
the decay rate of interest:
\beq
    \Ga_{\rm SM} (Z \rightarrow \bar{\nu}\nu )
    \, = \, \frac{g_z^2 m_Z}{96 \pi} \, \simeq \, 167 \, {\rm MeV}    \, .
    \label{5-16}
\eeq
Our result (\ref{5-12}) differs from the above by factor of
$\frac{2}{3} \left( \frac{m_\nu}{m_Z} \right)^2$, {\it i.e.},
\beq
    \Ga_{\rm SM} (Z \rightarrow \bar{\nu}\nu )
    \, \simeq \,
    \frac{2}{3} \left( \frac{m_\nu}{m_Z} \right)^2 \,
    \Ga_{hol} (Z \rightarrow \bar{\nu}\nu ) \, .
    \label{5-17}
\eeq
In the rest of this section we consider how such a discrepancy arises
and shall present explanations of the factors $\left( \frac{m_\nu}{m_Z} \right)^2$
and $\frac{2}{3}$, respectively.

\noindent
\underline{Nature of the massive holonomy operator}

Remember that the factor $\frac{1}{(12)(23)(31)}$ in (\ref{5-5})
and (\ref{4-23}) originates from the massive holonomy operator (\ref{4-21}).
The massive holomomy operator generates amplitudes of gluons (or massless gauge bosons
in general) and a pair of massive fermions.
We have naively applied the $\xi \zt$-prescription to the gluons in the above calculations
but in the holonomy formalism fermions and gluons are interpreted as superpartners
to each other. Hence, in principle, we need to use the identical mass for
massive deformations of both fermions and gluons.
This means that when we apply the massive holonomy operator to the calculation
of $Z$-boson decay processes $Z \rightarrow \bar{f}f$ the square of $\frac{1}{(12)(23)(31)}$
should be scaled as
\beq
    \left| \frac{1}{(12)(23)(31)} \right|^2 \, \longrightarrow \,
    \left( \frac{m_f}{m_Z} \right)^2 \, \left| \frac{1}{(12)(23)(31)} \right|^2 \, .
    \label{5-18}
\eeq
This prescription explains the factor of $\left( \frac{m_\nu}{m_Z} \right)^2$ in
(\ref{5-17}).
Since the massive holonomy operator itself does not contain information
of the massive gauge boson, it is reasonable to use the prescription (\ref{5-18})
and we may regard this as a principle in the decay-rate calculation
of the massive vector bosons in general.

From (\ref{5-1}) to (\ref{5-12}) we have considered the decay of
the $Z$-boson into a pair of neutrinos.
In this particular case, as mentioned earlier, the nonzero amplitudes
are given solely by the spin +1 state of the $Z$-boson, {\it i.e.},
\beq
    \Ga_{hol} (Z \rightarrow \bar{\nu}\nu ) \, = \,
    \Ga_{hol} (Z^{(+)} \rightarrow \bar{\nu}\nu ) \, .
    \label{5-19}
\eeq
For other types of decays this relation does not necessarily hold.
For example, in the case of $Z \rightarrow \bar{e} e$, we can similarly calculate
the decay rate as
\beqar
    \Ga_{hol} (Z \rightarrow \bar{e}e )
    & = &
    \Ga_{hol} (Z^{(+)} \rightarrow \bar{e}e )
    \, + \,
    \Ga_{hol} (Z^{(0)} \rightarrow \bar{e}e )
    \label{5-20} \\
    \Ga_{hol} (Z^{(+)} \rightarrow \bar{e}e )
    & = &
    \frac{ a_e^2 + b_e^2 }{ a_\nu^2 }
    \, \Ga_{hol} (Z^{(+)} \rightarrow \bar{\nu}\nu )  \, ,
    \label{5-21} \\
    \Ga_{hol} (Z^{(0)} \rightarrow \bar{e}e )
    & = &
    \frac{ b_e^2 }{ a_\nu^2 } \,
    \Ga_{hol} (Z^{(+)} \rightarrow \bar{\nu}\nu ) \, .
    \label{5-22}
\eeqar
where
\beq
    a_\nu \, = \, \hf \, , ~~~
    a_e \, = \, - \hf + \sin^2 \th_W \, , ~~~
    b_e = \sin^2 \th_W \, .
    \label{5-23}
\eeq
Thus, using the prescription (\ref{5-18}), we find that these results lead to the relation
\beq
    \Ga_{\rm SM} (Z \rightarrow \bar{e}e )
    \, \simeq \,
    \frac{2}{3} \left( \frac{m_e}{m_Z} \right)^2 \,
    \Ga_{hol} (Z^{(+)} \rightarrow \bar{e}e ) \, .
    \label{5-24}
\eeq
With the experimentally determined value
$\sin^2 \th_W \simeq 0.23$, we have $\frac{ a_e^2 + b_e^2 }{ a_\nu^2 } \simeq 0.50$.
Thus, taking account of (\ref{5-16}), we find that the relation (\ref{5-24})
indeed agrees with the experimental value,
$\Ga \left( Z \rightarrow \bar{e} e \right)  \simeq  84$ MeV.

In actuality, one can generalize the above results to the decay of
$Z \rightarrow \bar{f}f$.
To be explicit, the decay rate of the standard model (\ref{5-14}) can be related to
that of the holonomy formalism by
\beq
    \Ga_{\rm SM} (Z \rightarrow \bar{f}f )
    \, \simeq \,
    \frac{2}{3} \left( \frac{m_f}{m_Z} \right)^2 \,
    \Ga_{hol} (Z^{(+)} \rightarrow \bar{f}f )
    \label{5-25}
\eeq
with the supplemental relations
\beqar
    \Ga_{hol} (Z^{(-)} \rightarrow \bar{f}f ) &=& 0 \, ,
    \label{5-26} \\
    \Ga_{hol} (Z^{(0)} \rightarrow \bar{f}f ) &=&
    \frac{ b_f^2 }{ a_\nu^2 } \,
    \Ga_{hol} (Z^{(+)} \rightarrow \bar{f}f )
    \label{5-27}
\eeqar
where $a_\nu = \hf$ and $b_f = -Q_f \sin^2 \th_W$.

\noindent
\underline{Higgs-like scalars in the holonomy formalism}

These results illustrate that in the holonomy formalism the
spin-0 $Z$-boson needs to be treated differently.
It is intriguing to see that the massive scalar particles (\ref{3-17})
and the spin-0 massive gauge bosons (\ref{3-30})
share the same structure in terms of the Grassmann variables $\xi^\al$ and $\zt^\al$.
This implies, at least naively, that the Higgs-like scalar particle
may be described by a linear combination of ${\rm w}^{(0)\pm}( \xi ,\zt)$ and
${\rm z}^{(0)}( \xi ,\zt )$, bearing in mind that the Higgs-like scalar is neutral in charge.
Then the Yukawa-type terms in (\ref{4-10}), (\ref{4-11}) and
(\ref{4-14}) suggest that, in comparison to the standard electroweak, we
can regard ${\rm z}^{(0)}$ as the Higgs-like scalar up to some constant.
At present, it is not clear how ${\rm z}^{(0)}$ relates to the recently
confirmed massive scalar particle at 126 GeV.

As summarized in (\ref{5-25}), we can interpret that the full $Z$-boson decay rate can
be obtained by the decay rates of ${\rm z}^{(\pm)}$,
excluding the effect of ${\rm z}^{(0)}$, in the holonomy formalism.
In the spinor-helicity formalism the positive- and negative-spin states
are related to each other by taking the complex conjugates.
Thus, the spin $\pm 1$ states are taken into account in the computation of the
decay rate or the square of the decay amplitude.
Of course, the $Z$-boson should eventually be the spin-1 massive boson, having
three spin states $(\pm, 0)$.
Averaging the decay rate over these states and taking account of
the spin $\pm 1$ state contributions, we can then explain the factor
$\frac{2}{3}$ in (\ref{5-17})\footnote{
We have tried to find more reasonable arguments for the factor $\frac{2}{3}$ but
this explanation is the best we can think of at the present stage.
}.

\section{Concluding remarks}

In the present paper, we aim at formulating an electroweak model in the holonomy formalism.
We first review mass generation prescriptions for the description of
massive scalar particles and massive fermions in the holonomy formalism.
We then apply these prescriptions to the electroweak gauge bosons
and define the operators of the weak vector bosons, the photons and
the fundamental fermions as functions of the Grassmann variables.
The main results of this paper are given by the generating functional for
the electroweak interaction vertices in (\ref{4-23}) and the consequent
derivation of the $Z$-boson decay rates into a pair of fermions in (\ref{5-25}).

In the following we discuss limitations and (conceptual) advantages of our formalism
in comparison to the standard electroweak model.
We first emphasize that the holonomy formalism utilizes the spoinor-helicity
formalism in which the physical operators are labeled by the helicity/spin
and the numbering index while in the standard model the physical operators
are labeled by the helicity/spin and the Minkowski index (or the $\ga$-matrix index
for the current operators).
Apart from such fundamental differences, our definitions of the electroweak bosons (\ref{3-18}),
the interaction terms (\ref{4-1}) and the electroweak currents (\ref{4-4})-(\ref{4-7})
are the same as those of the standard electroweak model.
In fact, our motivation of this paper has been to seek for an alternative electroweak
model, starting from these standard definitions with introduction of
the coupling constants $( g , g^\prime)$, the Weinberg angle $\th_W$ and
the weak hypercharge ${\rm y}$.

\noindent
\underline{Lack of mass predictability}

In our formalism mass generation is essentially carried out
by massive deformation of the spinor-helicity formalism.
Technically, this is realized by off-shell continuation of Nair's superamplitude
method or what we name the $\xi \zt$-prescriptions.
This allows us to construct a Higgsless electroweak model and to define massive fermions
(\ref{2-14})-(\ref{2-17}) as well as the electroweak gauge bosons (\ref{3-30}), (\ref{3-31})
in terms of the Grassmann variables $\xi^\al$ and $\zt^\al$ ($\al = 1,2,3,4$) in (\ref{3-32}).
The essence of the massive deformation lies in the
definition of the ``massive'' Grassmann variable $\zt^\al = \frac{m}{(u \eta)}\eta^A \th^\al_A$
where $\eta^A$ is the reference spinor and $m$ denotes the mass of
the particle in question.
The reference spinors for fermions and weak bosons
can be chosen properly while masses are free input parameters in either case.
This means that, similarly to the standard model,
we can not theoretically predict masses of fermions or weak vector bosons
in our formalism.

\noindent
\underline{Peculiarity of the electroweak gauge group}

In the spinor-helicity formalism the Lorentz symmetry is given
by $SL (2 , {\bf C})_u \times SL (2 , {\bf C})_\bu$.
Physical operators are expressed in terms of the holomorphic spinor momenta,
satisfying half the Lorentz symmetry $SL (2 , {\bf C})_u = SU(2)_u^{\bf C}$.
Taking account of the $U(1)_u$ phase invariance of the holomorphic spinor momentum,
we find that the holomorphic part of the Loretnz symmetry includes the $U(2)_u$ group.
As mentioned in the beginning of section 3,
actions of the $U(2)_u$ transformations on the physical operators
are functionally equivalent to those of the $U(2)_{L {\rm y}}$
gauge transformations on the physical operators.
In this sense, the electroweak $U(2)_{L {\rm y}}$ gauge symmetry is special because
it would be interchangeable with the ``holomorphic'' Lorentz symmetry.
This means that we can generate mass of the gauge bosons by the Lorentz symmetry
breaking rather than the spontaneous breaking of the gauge symmetry.
Indeed, as discussed in section 3, the massive deformation of the spinor
momenta breaks the $U(2)_u$ symmetry down to $SU(2)_u \times U_{\hat u}$.

The $U(2)_{L {\rm y}}$ gauge symmetry is therefore the only symmetry that
is suitable for massive deformation of the gauge bosons via the
Lorentz symmetry breaking instead of the spontaneous gauge symmetry breaking.
This interpretation answers the question, ``Why do only the weak bosons
become massive in nature?'', and suggests conceptual and mathematical
motivations to construct a Higgsless model in the spinor-helicity formalism.

\noindent
\underline{Mathematical distinction between quarks and leptons}

In the holonomy formalism the coupling constant of an $SU(N)$ gauge boson
is determined by the reciprocal of the KZ parameter
$\kappa = k + h^{\vee}$ as shown in (\ref{3-8}).
According to Nair's original observation \cite{Nair:1988bq}, the level number
should be fixed at $k = 1$. The dual Coxeter number for $SU(N)$, on the other hand,
is given by $h^\vee = N$ in the highest weight representation (or the positive root system) and
$h^\vee = - N$ in the lowest weight representation (or the negative root system), respectively.
Remember that, by use of these relations, we assign the highest weight representation
of $SU(2)_L$ to the left-handed quark doublet and the lowest weight representation
of $SU(2)_L$ to the left-handed lepton doublet.
Thus our interpretation leads to natural distinction between quarks and leptons
in terms of a weight for the representation of $SU(2)_L$.

The quarks couple with gluons and form $SU(3)$ triplets while the leptons do not.
This fact then allows us to infer that the quarks, either left- or right-handed ones,
are also in the highest weight representation of $SU(3)$.
The corresponding coupling constant in the holonomy formalism is given by
\beq
    g_{hol} \, = \, \frac{1}{1+3} \, = \, \frac{1}{4} \, .
    \label{6-1}
\eeq
This factor should play a role similar to the weak hypercharge, however, obviously
the value 1/4 does not correspond to the conventional strong hypercharge.
This discrepancy can not be explained properly unless we integrate
QCD into the electroweak model; we shall investigate such a theory in a forthcoming paper.

\noindent
\underline{For three generations}

In the present paper we consider a one-generation model.
We thus neglect flavor mixing effects in the definition of the charged currents
(\ref{4-6}) and (\ref{4-7}).
To build a realistic model, however, we need to consider a model of three generations,
besides the incorporation of QCD.
In this paper we could not find any theoretical reasonings
for the origin of three generations; this question will be studied in the forthcoming paper.
It is, however, rather straightforward to write down the
charged currents in the three-generation model:
\beqar
    J^- &=& \bu_L^i \, (K^- )_{ij} \, d_L^j \, + \, {\bar \nu}_{L}^{i} \, (N^- )_{ij} \, e_{L}^{j}
    \, ,
    \label{6-2} \\
    J^+ &=&  \bd_L^i  \, (K^+ )_{ij} \, u_L^j \, + \, {\bar e}_{L}^{i} \, (N^+ )_{ij} \, \nu_{L}^{j}
    \label{6-3}
\eeqar
where $i, j = 1,2,3$ denote the generations.
$K^-$ and $N^+$ represent the Cabibbo-Kobayashi-Maskawa (CKM) matrix and
the Pontecorvo-Maki-Nakagawa-Sakata (PMNS) matrix, respectively.
$K^+$ and $N^-$ are obtained by the Hermite conjugates of these matrices, respectively,
{\it i.e.},
\beq
    \begin{array}{ll}
      K^- \, = \, (K^+ )^\dagger & ~ \mbox{: CKM matrix} \, , \\
      N^+ \, = \, (N^- )^\dagger & ~ \mbox{: PMNS matrix} \, .
    \end{array}
    \label{6-4}
\eeq

\noindent
\underline{Selection rules for the electroweak interactions}

The generating functional for the electroweak interaction vertices in (\ref{4-23})
originates from the massive holonomy operator (\ref{4-21})
which has been used to derive the generating functional for the massive fermion amplitudes.
As a consequence, (\ref{4-23}) deals only with those interactions that involve a pair of fermions.
There exist, however, other types of interactions, {\it e.g.},
couplings among the electroweak gauge bosons by themselves.
Since in a pure-boson system we can not restrict the number of interacting
particles, a generating functional for such interactions can not be
reduced in a form of (\ref{4-23}). It should be obtained more
directly from the massive holonomy operator, in this case
the massive holonomy operator used for the computation of the massive
scalar amplitudes \cite{Abe:2012en}. We have not constructed such
a generating functional in the present paper, focusing rather on
typical electroweak interactions, {\it i.e.}, the
couplings of the weak vector bosons with the charged and neutral currents.

In the holonomy formalism one can, however, easily tell the possible forms of
pure gauge-boson interactions by counting the total
number of Grassmann variables in the interactions of interest.
The number can be enumerated from the definitions of
the $W^\pm / Z$ boson operators (\ref{3-30}) and the photon operators (\ref{3-31}).
As discussed several times in this paper, the interactions vanish unless the number is eight
due to the Grassmann integrals over $\th_A^\al$.
In this sense the Grassmann integrals provide us selection rules for the electroweak interactions.

For example, since the minimum number of the Grassmann variables in the $W^\pm$-boson operators
is two, the number of $W^\pm$-boson legs in the tree-level pure $W^\pm$-boson interactions
is at most four.
This fact is in accord with the calculations in the standard electroweak model.
Similar analyses can be made for pure $Z$-boson interactions and  $W^\pm / Z$-$\ga$ interactions
in general.
It is intriguing that we can obtain these results
without taking low energy limits of the involving particles.

\vspace{0.2cm}


\end{document}